\documentstyle[aps,preprint]{revtex}
\begin{document}
\draft
\preprint{\vbox{\hbox{{\tt [ SOGANG-HEP 297/02 | hep-th/0207237]}}}}
\title{Exact self-gravitating $N$-body motion in the CGHS model}
\author
{Won Tae Kim\footnote{electronic address:wtkim@ccs.sogang.ac.kr} and
  Edwin J. Son\footnote{electronic address:sopp@string.sogang.ac.kr}}
\address{Department of Physics and Basic Science Research Institute,\\
         Sogang University, C.P.O. Box 1142, Seoul 100-611, Korea}
\date{\today}
\maketitle
\begin{abstract}
In the asymptotically flat two-dimensional dilaton gravity, we
present an $N$-body particle action which has a dilaton coupled mass
term for the exact solubility. This gives nonperturbative
exact solutions for the $N$-body self-gravitating system, so the
infalling particles form a black hole and their trajectories are
exactly described. In our two-dimensional case, the critical mass
for the formation of black holes does not exist, so even a single
particle forms a black hole, which means that we can treat many
black holes. The infalling particles give additional time-like
singularities in addition to the space-like black hole singularity.
However, the latter singularities can be properly cloaked by the
future horizons within some conditions.
\end{abstract}
\bigskip
There has been much interests in two-dimensional dilaton gravities
\cite{cghs} as toy models or a spherical symmetric reduction of
some higher-dimensional gravities because they have most of the
interesting properties of the four-dimensional gravity theories
even though they are much simpler than the original ones
\cite{rst,bc,gs}. In these works, the conformal fields play an
important role for the exact solubility. On the other hand, if
point particles as a matter source are considered, then it would
be worthwhile for studying the galactic evolution and star systems
\cite{my}. Recently, the $N$-body self-gravitating system in the
Jackiw-Teitelboim(JT) \cite{jt} gravity has been studied and the
closed solutions are obtained \cite{mro}, furthermore, the
explicit expressions are derived for the two equal mass
particles. In this case, the spacetime outside the moving matter
is not flat, in which the scalar curvature is constant.

So, one may think that the $N$-particle
self-gravitating system for which the asymptotic spacetime is
flat. It would be interesting to obtain nonperturbative exact
solutions for this case, which is not yet known. At first sight,
it seems to be easily realized by the
Callan-Giddings-Harvey-Strominger (CGHS) model \cite{cghs} coupled
to the $N$-particles. However, obtaining exact solutions is
generically difficult for the massive particles.

In this paper, we are concerned with a slightly modified particle
action to obtain the exact soluble model, which gives a remarkable
simple $N$-body self-gravitating solutions in the CGHS model. The
infalling particles form a black hole in the latest time and their
trajectories are exactly obtained. In the CGHS model, the critical
mass \cite{rst,ps,st} for the formation of black holes does not
exist, which means that the point particles themselves become
black holes although the infalling mass is even small. Therefore,
in our case, a single particle with a time-like curvature
singularity becomes a single black hole with a space-like
curvature singularity as time goes on. Intriguingly, if we
consider $N$-infalling massive particles, then the particles
become $N$-black holes, eventually, the combined holes appear as a
larger black hole. Finally, the cosmic sensorship is still valid
by the future horizons within some conditions.

We begin with the dilaton gravity action \cite{cghs} with the following
$N$-massive point particle action in two dimensions,
\begin{eqnarray}
\label{action} S &=& S_{DG} + S_P, \\
S_{DG} &=& \frac{1}{2\pi} \int d^2 x \sqrt{-g} e^{-2\phi}
  \left[ R + 4(\nabla\phi)^2 + 4\lambda^2 \right], \\
S_P &=& \frac{1}{2} \sum_{a=1}^N \int d^2 x \int d\tau_a
  \delta^2 ( x - z_a ( \tau_a ) ) e_a ( \tau_a ) \left[ e_a^{-2} ( \tau_a )
  g_{\mu\nu} (x) \frac{dz_a^\mu}{d\tau_a} \frac{dz_a^\nu}{d\tau_a}
  - m_a^2 e^{-2\alpha\phi(x)} \right],
\end{eqnarray}
where $g$ and $\phi$ are the metric and the dilaton field, and $\lambda^2$
is a cosmological constant. Note that $e_a$ and $z_a$ are the einbeins
and the coordinates for the $N$-particles, respectively.
The action of the original massive particle corresponding to $\alpha=0$
will be modified to obtain the exact geometry and the trajectories of
the massive particles.
For $\alpha=0$, in the fixed background geometry, the particle trajectory was
studied in Ref. \cite{hkp}. However, in this case, the self-gravitating
system was not realized because of some lack of global symmetry for
the massive particles.
From the previous work \cite{kk}, the kink coupled to the
two-dimensional dilaton gravity has been exactly solved, so we now
take $\alpha=1$ for the exact solubility in our particle case.

From the action (\ref{action}), the equations of motion for
the metric, dilaton, einbeins, and the coordinates are given as
\begin{eqnarray}
\label{em:gravity} & & 2e^{-2\phi} \left[ \nabla_\mu \nabla_\nu \phi
  - g_{\mu\nu} \left( \Box \phi - (\nabla\phi)^2
  + \lambda^2 \right) \right] = T_{\mu\nu}^P, \\
\label{em:dilaton} & & e^{-2\phi} \left[ R - 4(\nabla\phi)^2 + 4 \Box \phi
  + 4 \lambda^2 \right] - \frac{\pi}{\sqrt{-g}} \sum_{a} \int d\tau_a
  \delta^2 ( x - z_a ) e_a m_a^2 e^{-2\phi} = 0, \\
\label{em:einbein} & & e_a^{-2} g_{\mu\nu}(z_a) \frac{dz_a^\mu}{d\tau_a}
  \frac{dz_a^\nu}{d\tau_a} + m_a^2 e^{-2\phi(z_a)} = 0, \\
\label{em:geodesic} & & g_{\mu\nu}(z_a) \frac{d}{d\tau_a} \left( e_a^{-1}
  \frac{dz_a^{\nu}}{d\tau_a} \right) + \Gamma_{\mu\alpha\beta}(z_a) e_a^{-1}
  \frac{dz_a^\alpha}{d\tau_a} \frac{dz_a^\beta}{d\tau_a}
  - m_a^2 e_a e^{-2\phi(z_a)} \frac{\partial\phi}{\partial z_a^\mu} = 0,
\end{eqnarray}
respectively,
where the energy-momentum tensors due to the point masses are written as
\begin{equation}
\label{EMtensor} T_{\mu\nu}^P = \frac{\pi}{\sqrt{-g}} \sum_a \int d\tau_a
  \delta^2 (x-z_a) e_a^{-1} g_{\mu\alpha} g_{\nu\beta}
  \frac{dz_a^\alpha}{d\tau_a} \frac{dz_a^\beta}{d\tau_a}.
\end{equation}
Note that the $N$-particles are explicitly labelled as $a=1, 2,
\cdots, N$.

Combining eqs. (\ref{em:gravity}) and (\ref{em:dilaton}) yields
\begin{eqnarray}
\label{eq:reduced} e^{-2\phi} \left[ R + 2 \Box \phi \right] &=&
  \frac{\pi}{\sqrt{-g}} \sum_a \int d\tau_a \delta^2 (x-z_a) e_a
  \left[ e_a^{-2} g_{\mu\nu} \frac{dz_a^\mu}{d\tau_a} \frac{dz_a^\nu}{d\tau_a}
  + m_a^2 e^{-2\phi} \right] \nonumber \\
&=& 0,
\end{eqnarray}
which nicely vanishes by using the einbein equations of motion
(\ref{em:einbein}). It is crucial to obtain the exact geometry and
particle trajectories without any approximations.

In the conformal gauge defined by $g_{+-} = - \frac{1}{2} e^{2\rho}, g_{--} =
g_{++} = 0$, where $x^\pm = (x^0 \pm x^1)$, the above equations of
motion and the constraints are written as
\begin{eqnarray}
& & \label{conf:geo} 2 e^{-2\phi} \left[ 2 \partial_+ \phi \partial_- \phi
  - \partial_+ \partial_- \phi + \frac{1}{2} \lambda^2 e^{2\rho} \right]
  = T_{+-}^P, \\
& & \label{conf:dilaton} 8 e^{-2(\rho+\phi)} \left[ \partial_+ \partial_- \rho
  + 2 \partial_+ \phi \partial_- \phi - 2 \partial_+ \partial_- \phi
  + \frac{1}{2} \lambda^2 e^{2\rho} \right] \nonumber \\
& & \qquad - 2 \pi e^{-2\rho} \sum_a \int d\tau_a \delta^2 (x - z_a) e_a m_a^2
  e^{-2\phi} = 0, \\
& & \label{conf:einbein} e_a^{-2} e^{2\rho(z_a)} \frac{dz_a^+}{d\tau_a}
  \frac{dz_a^-}{d\tau_a} - m_a^2 e^{-2\phi(z_a)} = 0, \\
& & \label{conf:geodesic} \frac{d}{d\tau_a} \left( e_a^{-1}
  \frac{dz_a^\pm}{d\tau_a} \right) + 2 e_a^{-1}
  \frac{\partial \rho(z_a)}{\partial z_a^\pm} \frac{dz_a^\pm}{d\tau_a}
  \frac{dz_a^\pm}{d\tau_a} + 2 m_a^2 e_a e^{-2(\rho + \phi)(z_a)}
  \frac{\partial \phi}{\partial z_a^\mp} =0,
\end{eqnarray}
and
\begin{equation}
\label{conf:cons} 2 e^{-2\phi} \left[ \partial_\pm \partial_\pm \phi
  - 2 \partial_\pm \rho \partial_\pm \phi \right] = T_{\pm\pm}^P,
\end{equation}
where the energy-momentum tensors are
\begin{eqnarray}
\label{conf:emt1} T_{\pm\pm}^P &=& 2 \pi e^{-2\rho} \sum_a \int d\tau_a \delta^2
  (x - z_a) e_a^{-1} \frac{e^{4\rho}}{4} \frac{dz_a^\mp}{d\tau_a}
  \frac{dz_a^\mp}{d\tau_a}, \\
\label{conf:emt2} T_{+-}^P &=& 2 \pi e^{-2\rho} \sum_a \int d\tau_a \delta^2
  (x - z_a) e_a^{-1} \frac{e^{4\rho}}{4} \frac{dz_a^+}{d\tau_a}
  \frac{dz_a^-}{d\tau_a}.
\end{eqnarray}
Note that eq. (\ref{conf:emt2}) shows that the matter is no more
conformal,
while for the massless case it vanishes with the help of eq.
(\ref{conf:einbein}).

The key ingredient of the exact solubility is due to eq. (\ref{eq:reduced})
written as in the conformal gauge
\begin{equation}
\partial_+ \partial_- (\rho-\phi)=0,
\end{equation}
and then the residual symmetry can be fixed by choosing $\rho = \phi$ in the
Kruskal gauge.

For the sake of convenience, we now reparametrize as $m_a e_a d\tau_a =
d\lambda_a$, then eqs. (\ref{conf:einbein}) and (\ref{conf:geodesic}) become
\begin{equation}
\label{einbein} \frac{dz_a^+}{d\lambda_a}
  \frac{dz_a^-}{d\lambda_a} - e^{-4\rho(z_a)} = 0
\end{equation}
and
\begin{equation}
\label{geodesic} \frac{d^2z_a^\pm}{d\lambda_a^2} + 2
  \frac{\partial\rho(z_a)}{\partial z_a^\pm} \frac{dz_a^\pm}{d\lambda_a}
  \frac{dz_a^\pm}{d\lambda_a} + 2 e^{-4\rho(z_a)}
  \frac{\partial\rho}{\partial z_a^\mp} = 0,
\end{equation}
in the Kruskal gauge, respectively.

Then we get the following first order differential equation by integrating eq.
(\ref{geodesic}) by using eq. (\ref{einbein}),
\begin{equation}
\label{eq:particlem} \frac{dz_a^\pm}{d\lambda_a} = A_a^{(\pm)} e^{-2\rho(z_a)},
\end{equation}
where $A_a^{(\pm)}$ are integration constants, and we choose $A_a^{(\pm)}
> 0$ to make $z_a^\pm$ be increasing functions with respect to $\lambda_a$.
Furthermore, substituting eq. (\ref{eq:particlem}) into eq. (\ref{einbein}), we
get $A_a^{(+)} A_a^{(-)} = 1$. Simply rewriting the relation
(\ref{eq:particlem}) between $z_a^+$ and $z_a^-$ as
\begin{equation}
\frac{1}{A_a^{(+)}} \frac{dz_a^+}{d\lambda_a}
  = \frac{1}{A_a^{(-)}} \frac{dz_a^-}{d\lambda_a} = e^{-2\rho(z_a)},
\end{equation}
the trajectories of the particles are given as straight lines,
\begin{equation}
\label{rel:particle} z_a^+ = (A_a^{(+)})^2 \left( z_a^- + B_a \right),
\end{equation}
where $B_a$'s are taken as positive integration constants since
we shall consider only the spacetime in the
region of $x^+>0$ and $x^-<0$.

Now we explicitly calculate the energy-momentum tensors
(\ref{conf:emt1}) and
(\ref{conf:emt2})
by integrating with respect to $\tau_a$. After
reparametrizing $\tau_a$ as $\lambda_a$, $T_{++}^P$ is
integrated as
\begin{eqnarray}
T_{++}^P &=& \frac{\pi}{2} e^{-2\rho(x)} \sum_a m_a
  \int d\lambda_a \delta (x^+ - z_a^+(\lambda_a))
  \delta (x^- - z_a^-(\lambda_a)) e^{4\rho(z_a)} \left(
  \frac{dz_a^-}{d\lambda_a} \right)^2 \nonumber \\
&=& \frac{\pi}{2} \sum_a \int d\lambda_a \delta (x^+ - z_a^+(\lambda_a))
  F_a^{(+)} (\lambda_a) \nonumber \\
&=& \frac{\pi}{2} \sum_a \left. \frac{1}{| dz_a^+ / d\lambda_a |} F_a^{(+)}
  (\lambda_a) \right|_{x^+ = z_a^+(\lambda_a^{(+)})},
\end{eqnarray}
where $F_a^{(+)}= m_a \delta (x^- - z_a^-(\lambda_a)) e^{2\rho(z_a)} \left(
\frac{dz_a^-}{d\lambda_a} \right)^2$.

Next by using eqs. (\ref{eq:particlem}) and (\ref{rel:particle}),
it is explicitly calculated as
\begin{eqnarray}
\label{int:emt1} T_{++}^P
&=& \frac{\pi}{2} \sum_a m_a \frac{(A_a^{(-)})^2}{A_a^{(+)}}
  \delta (x^- - z_a^-(\lambda_a^{(+)})) \nonumber \\
&=& \frac{\pi}{2} \sum_a m_a \frac{1}{(A_a^{(+)})^3}
  \delta \left( x^- - \frac{z_a^+(\lambda_a^{(+)})}{(A_a^{(+)})^2} + B_a
  \right) \nonumber \\
&=& \frac{\pi}{2} \sum_a m_a \frac{1}{(A_a^{(+)})^3}
  \delta \left( \frac{x^+}{(A_a^{(+)})^2} - x^- - B_a \right).
\end{eqnarray}
Note that the relation $x^+ = z_a^+(\lambda_a^{(+)})$ was used.

On the other hand, one might wonder if we integrate
$\delta(x^- - z_a^-)$ instead of $\delta(x^+ - z_a^+)$ in the first line
of eq. (\ref{int:emt1}), then the
similar calculation yields the same result,
\begin{eqnarray}
\label{int:emt2} T_{++}^P &=& \frac{\pi}{2} \sum_a \int d\lambda_a \delta (x^- -
  z_a^-(\lambda_a)) F_a^{(-)} (\lambda_a) \nonumber \\
&=& \frac{\pi}{2} \sum_a \left. \frac{1}{| dz_a^- / d\lambda_a |} F_a^{(-)}
  (\lambda_a) \right|_{x^- = z_a^-(\lambda_a^{(-)})}
  \nonumber \\
&=& \frac{\pi}{2} \sum_a m_a \frac{1}{(A_a^{(+)})^3} \delta \left(
  \frac{x^+}{(A_a^{(+)})^2} - x^- - B_a \right),
\end{eqnarray}
where $F_a^{(-)}= m_a \delta (x^+ - z_a^+(\lambda_a)) e^{2\rho(z_a)} \left(
\frac{dz_a^-}{\lambda_a} \right)^2$.
Here, $\lambda_a^{(\pm)}$'s satisfy
\begin{equation}
\label{cond:lambda} z_a^\pm ( \lambda_a^{(\pm)} ) = x^\pm.
\end{equation}
Therefore, the two expressions (\ref{int:emt1}) and (\ref{int:emt2}) derived
from different integration steps are coincident.

Similarly, the other energy-momentum
tensors in the Kruskal gauge can be obtained in this way, which are compactly
written as
\begin{eqnarray}
\label{EM++} T_{++}^P &=& \frac{\pi}{2} \sum_a m_a \frac{1}{(A_a^{(+)})^3}
  \delta \left( \frac{x^+}{(A_a^{(+)})^2} - x^- - B_a \right), \\
\label{EM--} T_{--}^P &=& \frac{\pi}{2} \sum_a m_a A_a^{(+)}
  \delta \left( \frac{x^+}{(A_a^{(+)})^2} - x^- - B_a \right), \\
\label{EM+-} T_{+-}^P &=& \frac{\pi}{2} \sum_a \frac{m_a}{A_a^{(+)}}
  \delta \left( \frac{x^+}{(A_a^{(+)})^2} - x^- - B_a \right).
\end{eqnarray}

Now integrating eq. (\ref{conf:geo}) with the energy-momentum tensor
(\ref{EM+-}), we obtain the metric solution,
\begin{eqnarray}
e^{-2\rho} &=& - \lambda^2 x^+ x^- + a_+(x^+) + a_-(x^-) \nonumber \\
& & - \frac{\pi}{2} \sum_a m_a A_a^{(+)} \left( \frac{x^+}{(A_a^{(+)})^2} - x^-
  - B_a \right) \theta \left( \frac{x^+}{(A_a^{(+)})^2} - x^- - B_a \right),
\end{eqnarray}
where $a_\pm(x^\pm)$ are integration functions which are
determined by the constraints (\ref{conf:cons}),
\begin{equation}
\label{eq:a} \partial_\pm \partial_\pm a_\pm (x^\pm) = 0.
\end{equation}
Note that the functions $a_\pm(x^\pm)$ in our
case are simply harmonic, while for the case of the conformal fields
they are determined by the conformal sources \cite{cghs}.

Integrating eq. (\ref{eq:a}), we obtain the solution of geometry for
the case of massive particles,
\begin{eqnarray}
\label{geometry} e^{-2\rho} &=& - \lambda^2 x^+ x^- + \lambda C_- x^+ + \lambda
  C_+ x^- + D \nonumber \\
& & - \frac{\pi}{2} \sum_a m_a A_a^{(+)} \left( \frac{x^+}{(A_a^{(+)})^2} - x^-
  - B_a \right) \theta \left( \frac{x^+}{(A_a^{(+)})^2} - x^- - B_a \right)
  \nonumber \\
&=& - \lambda^2 \left( x^+ - \frac{h^{(+)}}{\lambda} \right) \left( x^- +
  \frac{h^{(-)}}{\lambda} \right) + \frac{M}{\lambda},
\end{eqnarray}
where $C_+$, $C_-$, and $D$ are integration constants, and
$h^{(+)} = C_+ +
\frac{\pi}{2\lambda} \sum_a m_a A_a^{(+)} \theta (\frac{x^+}{(A_a^{(+)})^2} -
x^- - B_a)$, $h^{(-)} = - C_- + \frac{\pi}{2\lambda} \sum_a
\frac{m_a}{A_a^{(+)}} \theta (\frac{x^+}{(A_a^{(+)})^2} - x^- - B_a)$, and $M =
\lambda D - \lambda h^{(+)} h^{(-)} + \frac{\lambda\pi}{2} \sum_a m_a A_a^{(+)}
B_a \theta (\frac{x^+}{(A_a^{(+)})^2} - x^- - B_a)$, and $\theta (x)$ is 1 for
$x>0$ and 0 for $x<0$. Then we require the linear
dilaton vacuum (LDV) boundary condition in the region that $x^+ \to 0^+$ and
$x^- \to 0^-$ where there are no particles.
Then the constants $C_\pm$ and $D$ are fixed to zero.
From the condition of $\partial_+ \rho = 0$, the
horizon curve is obtained as
\begin{equation}
\label{horizon} - \lambda x^- = h^{(-)}.
\end{equation}

Now the mass function $M$ is defined by
\begin{eqnarray}
M &=& \frac{\lambda\pi}{2} \sum_a m_a A_a^{(+)} B_a \theta \left(
  \frac{x^+}{(A_a^{(+)})^2} - x^- - B_a \right) - \lambda h^{(+)} h^{(-)}
  \nonumber \\
&=& \frac{\lambda\pi}{2} \sum_a m_a A_a^{(+)} \theta \left(
  \frac{x^+}{(A_a^{(+)})^2} - x^- - B_a \right) \nonumber \\
& & \qquad \qquad \qquad \times \left[ B_a -
  \frac{\pi}{2\lambda^2} \sum_b \frac{m_b}{A_b^{(+)}} \theta \left(
  \frac{x^+}{(A_b^{(+)})^2} - x^- - B_b \right) \right].
\end{eqnarray}
The mass $M$ increases as time goes on such that in the latest time
the total mass $M_T$ becomes
\begin{equation}
\label{mass:tot} M_T = \frac{\lambda\pi}{2} \sum_a m_a A_a^{(+)} \left[ B_a
  - \frac{\pi}{2\lambda^2} \sum_b \frac{m_b}{A_b^{(+)}} \right],
\end{equation}
where the following condition is required to make $M_T$ positive definite,
\begin{equation}
\label{cond:positiveM} \sum_a m_a A_a^{(+)} B_a > \frac{\pi}{2\lambda^2}
  \sum_{a,b} m_a m_b \frac{A_a^{(+)}}{A_b^{(+)}}.
\end{equation}

To investigate the curvature singularities,
we now calculate the curvature scalar $R$,
\begin{eqnarray}
\label{curvature} R &=& 8 e^{-2\rho} \partial_+ \partial_- \rho \nonumber \\
&=& \frac{4\lambda M}{\frac{M}{\lambda} - \lambda^2 (x^+ -
  \frac{h^{(+)}}{\lambda}) (x^- + \frac{h^{(-)}}{\lambda})} - 2 \pi \sum_a
  \frac{m_a}{A_a^{(+)}} \delta \left( \frac{x^+}{(A_a^{(+)})^2} - x^- -B_a
  \right),
\end{eqnarray}
which reads curvature singularity curves,
\begin{eqnarray}
\label{eq:singularity1} \lambda^2 \left( x^+ - \frac{h^{(+)}}{\lambda} \right)
  \left( x^- + \frac{h^{(-)}}{\lambda} \right) &=& \frac{M}{\lambda}, \\
\label{eq:singularity2} x^+ - \left( A_a^{(+)} \right)^2 \left( x^- + B_a
  \right) &=& 0.
\end{eqnarray}
Note that the infalling particles form the black hole which have
curvature singularity similarly to the original CGHS model coupled to
the conformal fields. In our case, the crucial difference appears
in the additional singularities which come from the particles
themselves. In this regard, the particles may be interpreted as
black holes since even a  single particle can collapse to the black hole
with the event horizon in our model, which is easily seen from
the above whole equations by simply setting $N=1$. The $N$-particles
eventually form a larger black hole.

Let us now check whether the singularity curves
(\ref{eq:singularity1})  and (\ref{eq:singularity2}) are naked or
not. The first curve (\ref{eq:singularity1}) from the curvature
singularities is naturally cloaked by the horizon (\ref{horizon}),
which satisfies the cosmic sensorship scenario. As for the other
time-like curve, from eqs. (\ref{eq:singularity2}) and (\ref{horizon}), we
can require $h^{(-)} > \lambda B_a$ in the latest time, in order
to make the singularity curve completely shielded by the horizon
curve (\ref{horizon}), {\it i.e.},
\begin{equation}
\label{cond:shielded} B_a < \frac{\pi}{2\lambda^2} \sum_b
  \frac{m_b}{A_b^{(+)}}.
\end{equation}
So all the singularities seem to be cloaked inside the horizon.
However, this condition (\ref{cond:shielded}) makes the total
mass (\ref{mass:tot}) be negative.
Therefore, as far as we require the total mass to be positive,
we find that the naked singularities partially appear. These naked
singularities are time-like compatible with the particle
trajectories.

Next, we obtain the particle trajectories in terms of the
parameter $\lambda_a$ by substituting eq.
(\ref{geometry}) into eq. (\ref{eq:particlem}),
\begin{eqnarray}
\frac{dz_a^-}{d\lambda_a} &=& A_a^{(-)} \left[ - \lambda^2 \left( z_a^+
  - \frac{h_a^{(+)}}{\lambda} \right) \left( z_a^- + \frac{h_a^{(-)}}{\lambda}
  \right) + \frac{M_a}{\lambda} \right] \nonumber \\
&=& - \lambda^2 A_a^{(+)} \left( z_a^- + B_a - \frac{h_a^{(+)}}{\lambda
  (A_a^{(+)})^2} \right) \left( z_a^- + \frac{h_a^{(-)}}{\lambda} \right) +
  A_a^{(-)} \frac{M_a}{\lambda},
\end{eqnarray}
where we used eq.\ (\ref{rel:particle}) and $h_a^{(+)} = \frac{\pi}{2\lambda}
\sum_b m_b A_b^{(+)} \theta \left( z_a^+ / (A_b^{(+)})^2 - z_a^- - B_b \right)$,
$h_a^{(-)} = \frac{\pi}{2\lambda} \sum_b \frac{m_b}{A_b^{(+)}} \theta \left(
z_a^+ / (A_b^{(+)})^2 - z_a^- - B_b \right)$, and $M_a = \frac{\lambda\pi}{2}
\sum_b m_b A_b^{(+)} B_b \theta \left( z_a^+ / (A_b^{(+)})^2 - z_a^- - B_b
\right) - \lambda h_a^{(+)} h_a^{(-)}$. Rearranging this equation, we get
\begin{eqnarray}
\label{de:particle} \frac{dz_a^-}{d\lambda_a} = - \lambda^2 A_a^{(+)} \left[
  \left( z_a^- + \zeta_a \right)^2 - \xi_a^2 \right],
\end{eqnarray}
where we define $\zeta_a = \frac{\lambda B_a - h_a^{(+)} /
(A_a^{(+)})^2 + h_a^{(-)}}{2\lambda}$ and $\xi_a^2 = \frac{1}{4}
\left( \frac{h_a^{(+)}}{\lambda (A_a^{(+)})^2} +
\frac{h_a^{(-)}}{\lambda} - B_a \right)^2 + \frac{M_a}{\lambda^3
(A_a^{(+)})^2}$, and take $\xi_a > 0$. We can divide the
worldline of a particular particle from worldlines of other
particles. In each region, we can think $h_a^{(+)}$, $h_a^{(-)}$,
and $M_a$ as constants, so $\zeta_a$ and $\xi_a$ also can be
thought as constants. We represent these as $\zeta_{ai}$ and $\xi_{ai}$,
where $i$ is an index for regions.

We now parametrize $z_a^- + \zeta_{ai} = \xi_{ai} \tanh \eta_{ai}$ for
$|z_a^- + \zeta_{ai}| < \xi_{ai}$ and $z_a^- + \zeta_{ai} = \xi_{ai}
\coth \eta_{ai}$ for $|z_a^- + \zeta_{ai}| > \xi_{ai}$.
First, if $|z_a^- + \zeta_{ai}| < \xi_{ai}$, then $dz_a^- = \xi_{ai}
{\rm sech}^2 \eta_{ai} d\eta_{ai}$, and eq. (\ref{de:particle}) is written as
\begin{equation}
\frac{\xi_{ai} {\rm sech}^2 \eta_{ai}}{\xi_{ai}^2 ( \tanh^2 \eta_{ai} - 1 )}
  \frac{d\eta_{ai}}{d\lambda_a} = - \lambda^2 A_a^{(+)}.
\end{equation}
Thus we obtain the solution of particle motion for $|z_a^- + \zeta_{ai}| <
\xi_{ai}$,
\begin{equation}
\label{sol1:particlem} z_a^- = \xi_{ai} \tanh \eta_{ai} - \zeta_{ai} = \xi_{ai} \tanh \left[
  \lambda^2 A_a^{(+)} \xi_{ai} (\lambda_a - \Lambda_{ai}) \right] - \zeta_{ai},
\end{equation}
where $\Lambda_{ai}$ is a constant of integration. Then from
(\ref{eq:particlem}), $z_a^+$ becomes
\begin{equation}
\label{sol2:particlem} z_a^+ = (A_a^{(+)})^2 \left[ \xi_{ai} \tanh \left(
  \lambda^2 A_a^{(+)} \xi_{ai} (\lambda_a - \Lambda_{ai}) \right) + B_a - \zeta_{ai}
  \right].
\end{equation}
Eqs. (\ref{sol1:particlem}) and (\ref{sol2:particlem}) are desirable increasing
functions with respect to $\lambda_a$.

Next, if $|z_a^- + \zeta_{ai}| > \xi_{ai}$, then $dz_a^- = - \xi_{ai}
{\rm csch}^2 \eta_{ai} d\eta_{ai}$, eq. (\ref{de:particle}) is written as
\begin{equation}
- \frac{\xi_{ai} {\rm csch}^2 \eta_{ai}}{\xi_{ai}^2 ( \coth^2 \eta_{ai} - 1 )}
  \frac{d\eta_{ai}}{d\lambda_a} = - \lambda^2 A_a^{(+)}.
\end{equation}
Thus we obtain the solution of particle motion for $|z_a^- + \zeta_{ai}| > \xi_{ai}$,
\begin{equation}
z_a^- = \xi_{ai} \coth \eta_{ai} - \zeta_{ai} = \xi_{ai} \coth \left[
  \lambda^2 A_a^{(+)} \xi_{ai} (\lambda_a - \Lambda_{ai}^{(1)}) \right] - \zeta_{ai},
\end{equation}
where $\Lambda_{ai}^{(1)}$ is a constant of integration. But this is not a
suitable solution, because it is a decreasing function with respect
to $\lambda_a$. Thus we require a restriction, $|z_a^- + \zeta_{ai}| < \xi_{ai}$.

We have considered CGHS model coupled to modified massive $N$-particles and
obtained the exact solutions.
The einbein formulation of the action (\ref{action}) gives a convenient
massless limit. This is a conformal particle case which is very
similar
to the original CGHS model. From now on, we shall present briefly the results
for the massless case.

We reparametrize $e_a d\tau_a = d\lambda_a$ for the massless case,
then eqs. (\ref{conf:einbein}) and (\ref{conf:geodesic}) become
\begin{equation}
\label{eq:einbein} e^{2\rho(z_a)} \frac{dz_a^+}{d\lambda_a}
  \frac{dz_a^-}{d\lambda_a} = 0
\end{equation}
and
\begin{equation}
\label{eq:geodesic} \frac{d^2z_a^\pm}{d\lambda_a^2} + 2
  \frac{\partial\rho(z_a)}{\partial z_a^\pm} \frac{dz_a^\pm}{d\lambda_a}
  \frac{dz_a^\pm}{d\lambda_a} = 0,
\end{equation}
in the Kruskal gauge, respectively.
As easily seen in eq. (\ref{eq:einbein}), one is an infalling case,
$\frac{dz_a^+}{d\lambda_a}=0$, and the other is outgoing one,
$\frac{dz_a^-}{d\lambda_a}=0$. However, we shall
consider only the ingoing mode as the infalling particles.

Integrating eq. (\ref{eq:geodesic}) gives
\begin{equation}
\label{eq:particle} \frac{dz_a^-}{d\lambda_a} = A_a e^{-2\rho(z_a)},
\end{equation}
where $A_a=e^{\rm const.}>0$ is a integration constant.

The energy-momentum tensors (\ref{conf:emt1}) and (\ref{conf:emt2})
are simply reduced to
\begin{eqnarray}
\label{eq:EMtensor1} T_{++}^P &=& \frac{\pi}{2} \sum_a A_a
  \delta(x^+ - z_a^+), \\
\label{eq:EMtensor2} T_{--}^P &=& 0 = T_{+-}^P,
\end{eqnarray}
where we used the relation (\ref{eq:particle}) and $z_a^+$'s are constants of
motion which are fixed by initial conditions.
Next, integrating the gravity equation (\ref{conf:geo}) and constraints
(\ref{conf:cons}) with the energy-momentum tensor (\ref{eq:EMtensor1}) and
(\ref{eq:EMtensor2}), the geometric solution up to
constant translations of $x$ is obtained as
\begin{eqnarray}
\label{sol:gravity} e^{-2\phi} &=& e^{-2\rho} \nonumber \\
&=& - \lambda^2 x^+ x^- - \frac{\pi}{2} \sum_a A_a (x^+ - z_a^+)
  \theta(x^+ - z_a^+) \nonumber \\
&=& - \lambda^2 x^+ \left( x^- + \frac{h^{(-)}}{\lambda} \right) +
  \frac{M}{\lambda},
\end{eqnarray}
by considering the boundary condition of linear dilaton vacuum
(LDV). Note that $h^{(-)} =
\frac{\pi}{2\lambda} \sum_a A_a \theta ( x^+ - z_a^+ )$, $M =
\frac{\lambda\pi}{2} \sum_a A_a z_a^+ \theta ( x^+ - z_a^+ )$
which is positive definite. After all particles collapsed in the
latest time, the total mass becomes $M_T = \frac{\lambda\pi}{2}
\sum_a A_a z_a^+$. The horizon curve is also defined as $ -
\lambda x^- = h^{(-)}$. On the other hand, the curvature is
written as $R = 4 \lambda M/ (\frac{M}{\lambda} - \lambda^2 x^+
  ( x^- + \frac{h^{(-)}}{\lambda} )) $
which reads curvature singularity curve,
$\lambda^2 x^+ \left( x^- + \frac{h^{(-)}}{\lambda}
  \right) = \frac{M}{\lambda}$.

If we consider only a single particle, {\it i.e.}, $N=1$, then the
metric solution, the horizon, and the mass formula are all the
same with those of the CGHS model since the infalling source
(\ref{eq:EMtensor1}) is the familiar delta-functional source in
the CGHS model. As for the difference of our model from the CGHS
model, the location of the matter is essentially dynamical, which
means that $z_a^+$ is a particle coordinate instead of a constant.
However, for the massless limit, the solution of the particle
coordinate is given as coincidentally a constant, so that the
whole structure is the same with that of the CGHS model.

Similar to the case of massive particles, we have many regions separated by
the worldlines of particles, $x^+ = z_a^+$. To make this explicit,
by combining eqs. (\ref{eq:particle}) and (\ref{sol:gravity}),
the geodesic equation becomes
\begin{equation}
\frac{dz_a^-}{d\lambda_a} = - \lambda^2 A_a  z_a^+ z_a^- - C_a,
\end{equation}
where $C_a = \frac{\pi}{2} \sum_b A_a A_b (z_a^+ - z_b^+) \theta (z_a^+ - z_b^+)
> 0$ is a constant. Integrating this equation, we obtain
\begin{equation}
| \lambda B_a z_a^- + C_a | = D_a e^{- \lambda B_a \lambda_a},
\end{equation}
where $B_a = A_a \lambda z_a^+ $ and $D_a = e^{\rm const.} $ are
constants. Then the particle solutions are explicitly written as
\begin{equation}
\label{sol:particle} - \lambda z_a^- = \left\{
\begin{array}{lcr}
\frac{D_a}{B_a} e^{-\lambda B_a\lambda_a} + \frac{C_a}{B_a}, &\qquad&
  {\rm for} \quad - \lambda z_a^- > \frac{C_a}{B_a} \\
- \frac{D_a}{B_a} e^{-\lambda B_a\lambda_a} + \frac{C_a}{B_a}, &\qquad&
  {\rm for} \quad - \lambda z_a^- < \frac{C_a}{B_a}
\end{array}
\right. ,
\end{equation}
where $z_a^+$ is constant and
 $B_a$, $C_a$, and $D_a$ are all positive constants.

In the original CGHS model, it has been well appreciated that an
infalling matter basically forms a black hole no matter how the
energy of the infalling matter is small. This fact might be more
or less unphysical, however, this feature enables us to study some
of characteristic features of many black holes as a merit in a
simple way. Of course, complicated collisions are still missing
because of the dimensional simplicity. In our model, the particles
can be interpreted as black holes and their motions are more or
less trivial in that their trajectories are straight line in the
Kruskal diagram. This triviality comes from the conformal factor
in front of the mass term, which plays an important role of the
exact solubility.

The final point to be mentioned is that the singularity of the
curvature scalar (\ref{curvature}) seems to be unusual, however,
it is in fact expected result because the classical particles are
regarded as points so that the particle radius is zero which is
much less than the horizon. Therefore, the particles have
curvature singularities from the classical point of view. In our
work, the classical particles evolves into the black holes in a
finite light cone time, and the corresponding time-like
singularities are changed into the space-like curvature
singularities. Of course, the time-like singularities partially
appear in our spacetime, while the space-like singularity is
completely cloaked by the event horizon. \vspace{1cm}

{\bf Acknowledgments}\\
This work was supported by the Korea Research Foundation Grant,
KRF-2001-015-DP0083.

\end{document}